\documentclass[aps,prl,floatfix,twocolumn,superscriptaddress]{revtex4}
\usepackage{latexsym}
\usepackage{graphicx}
\usepackage{rotating}
\usepackage{hyperref}
\usepackage{amsmath,amssymb,amsfonts}
\usepackage{bm}
\usepackage{color}

\newcommand{\dxy}{d$_{xy}$ }
\newcommand{\dyz}{d$_{yz}$ }
\newcommand{\dxz}{d$_{xz}$ }
\newcommand{\dzd}{d$_{3z^{2}-r^{2}}$}
\newcommand{\dxd}{d$_{x^{2}-y^{2}}$}

\newcommand{\dyzd}{d$_{yz}$}
\newcommand{\dxzd}{d$_{xz}$}
\newcommand{\vo}{VO$_2$ }
\newcommand{\vox}{VO$_2$}
\newcommand{\tg}{t$_{2g}$ }
\newcommand{\eg}{e$_{g}$ }
\newcommand{\ag}{a$_{1g}$ }

\newcommand{\emphasize}{\emph}

\def\onlinecite#1{\cite{#1}}

\begin{document}

\title{Vanadium dioxide : A Peierls-Mott insulator stable against disorder}
\author{C\'edric Weber}
\affiliation{Cavendish Laboratory,  J. J. Thomson Av., Cambridge CB3 0HE, U.K.}
\author{David D. O'Regan}
\affiliation{Cavendish Laboratory,  J. J. Thomson Av., Cambridge CB3 0HE, U.K.}
\affiliation{Theory and Simulation of Materials, \'{E}cole Polytechnique F\'{e}d\'{e}rale de Lausanne, Station 12, 1015 Lausanne, Switzerland}
\author{Nicholas D. M. Hine}
\affiliation{Cavendish Laboratory,  J. J. Thomson Av., Cambridge CB3 0HE, U.K.}
\affiliation{Department of Materials, Imperial College London, Exhibition Road, London SW7 2AZ, U.K.}
\author{Mike C. Payne}
\affiliation{Cavendish Laboratory,  J. J. Thomson Av., Cambridge CB3 0HE, U.K.}
\author{Gabriel Kotliar}
\affiliation{Rutgers University, 136 Frelinghuysen Road, Piscataway, NJ, U.S.A.}
\author{Peter B. Littlewood}
\affiliation{Cavendish Laboratory,  J. J. Thomson Av., Cambridge CB3 0HE, U.K.}
\affiliation{Physical Sciences and Engineering, Argonne National Laboratory, Argonne, Illinois 60439, U.S.A.}


\begin{abstract}
Vanadium dioxide undergoes a first order metal--insulator transition at 340 K.
In this work, we develop and carry out state of the art linear scaling DFT calculations refined with non-local dynamical mean-field theory.
We identify a complex mechanism, a Peierls-assisted orbital selection Mott instability, which is responsible for the insulating M$_1$ phase, and 
furthermore survives a moderate degree of disorder.
\end{abstract}

\maketitle


Vanadium dioxide (\vox) undergoes a first order metal--insulator transition (MIT) at 340 K \cite{vo2_paper_ref_MIT}.
At high temperature, the crystal structure is metallic with the rutile structure (R), while it transforms to the monoclinic (M$_1$) 
phase and becomes insulating below the transition temperature. 
The nature of the metal-insulator transition in \vo has been long discussed, with particular emphasis placed on the role of electron correlations in forming the charge gap.
Photoemission experiments give strong evidence for strong electron-electron and electron-phonon coupling in \vo \cite{vo2_paper_ref5}, suggesting that this compound is an archetypal Mott insulator. However, density functional theory (DFT) predicts 
the M$_1$ phase to be metallic \cite{vo2_paper_ref_8_evidence_strong_coulomb_renormalization,vo2_paper_ref1}. 
An alternative point of view is that the low-temperature phase of \vo may constitute a band (Peierls) insulator,
where the crystal distortion with the V-V dimerization
splits the \ag bonding band, as suggested early by Goodenough \cite{vo2_paper_ref_goodenough}. 
Lastly one should consider a charge transfer insulator exhibiting a strong mass renormalisation \cite{vo2_large_covalency_charge_transfer_k_edge}.
The purpose of this letter is to disentangle these competing pictures.
The Peierls picture was supported by DFT+GW calculations, where the authors found that off-diagonal matrix elements in the self-energy opened a gap \cite{vo2_paper_ref1}, although
its value was almost zero and thus well below the experimental value of $0.6$~eV \cite{vo2_paper_ref_gap}.
Very recently, a model Hamiltonian approach using cluster dynamical mean-field theory (DMFT) applied to a three band Hamiltonian for the \tg orbitals has been shown to successfully capture the insulating nature of the M$_1$ phase \cite{vo2_paper_ref11_dynamical_singlet,gabi_vo2}, and the authors found a charge gap of $0.6$~eV, in very good agreement with experiment \cite{vo2_paper_ref_gap}.
Hence, \vo is, in the latter view, not a conventional Mott insulator. Instead, the formation of dynamical V-V singlet pairs due to strong Coulomb correlations is necessary to trigger the opening of a Peierls gap.
We note, however, that in Ref.~\onlinecite{vo2_paper_ref11_dynamical_singlet} the vanadium 
3d subshell is occupied by a single electron ($0.8$ electrons for the \ag with only $0.1$ electrons remaining in each of the e$_{ \pi g}$ orbitals).
A general problem with model Hamiltonian approaches, recently pointed out in Ref.~\onlinecite{millis_double_counting_covalency}, is that the 3d orbital density is very much affected by the orbital subset projection used in the calculations.
In particular, it has been shown recently using XAS measurements  \cite{vo2_paper_ref2} that the states of \vo are not well characterized by a single dominant ionic configuration, rather exhibiting a distributed orbital character, suggesting room for correction of Goodenough's ionic picture of \vox.

The key issues that we address in this work are: 
(1) Is the 3d$^1$ ionic picture of Goodenough valid and how many electrons are involved in the orbital selection process;
(2) Can Mott correlations alone drive \vo to an insulator, and what is the minimal local repulsion $U_d$ necessary to localize the charge, i.e., 
the Zaanen-Sawatzky-Allen (ZSA) boundary \cite{zsa_paper,brinkman_correlated}; (3) How is the ZSA boundary affected by other localization processes, such as the
Anderson charge localization induced by disorder, and can we find an insulator for a combination of realistic disorder and Coulomb repulsion; 
(4) Are non-local corrections to the self energy (the Peierls mechanism) an essential ingredient
to trigger the gap opening for reasonable local repulsion $U_d$, and is the latter insulating phase stable against external perturbations such as disorder.
To address these points, we move beyond the model Hamiltonian approach and investigate the effect of correlations in a disordered prototype for the metal--insulator transition in \vo from 
the ab-initio perspective. We study the M$_1$ phase of \vo using first-principles calculations as a function of static disorder
with a state of the art linear scaling DFT method \cite{onetep_main_ref}. The capability of linear scaling DFT 
to describe large super-cells, containing several hundreds of atoms, is necessary to comprehensively 
tackle the issue of disorder. We extend our DFT calculations with the DMFT approximation
\cite{OLD_GABIS_REVIEW,dca_reference_cluster_dmft} in order to refine the description of the strong correlations induced
by the 3d subshell of the vanadium sites (for more details see the supplementary material). 
Throughout this work we used typical values for the screened Coulomb interaction ($U=4$~eV)
and HundÕs coupling ($J=0.68$~eV) \cite{vo2_paper_U,vo2_paper_J},
and our calculations were carried out for 324 atom super-cells (108 V atoms) and 768 atom super-cells (256 V atoms) at fixed temperature $T=189$~K. All orbitals are 
defined in the local coordinate system \cite{vo2_paper_ref_10_rotation_axis} associated with the vanadium atoms. 

\begin{figure}
\begin{center}
\includegraphics[width=1.0\columnwidth]{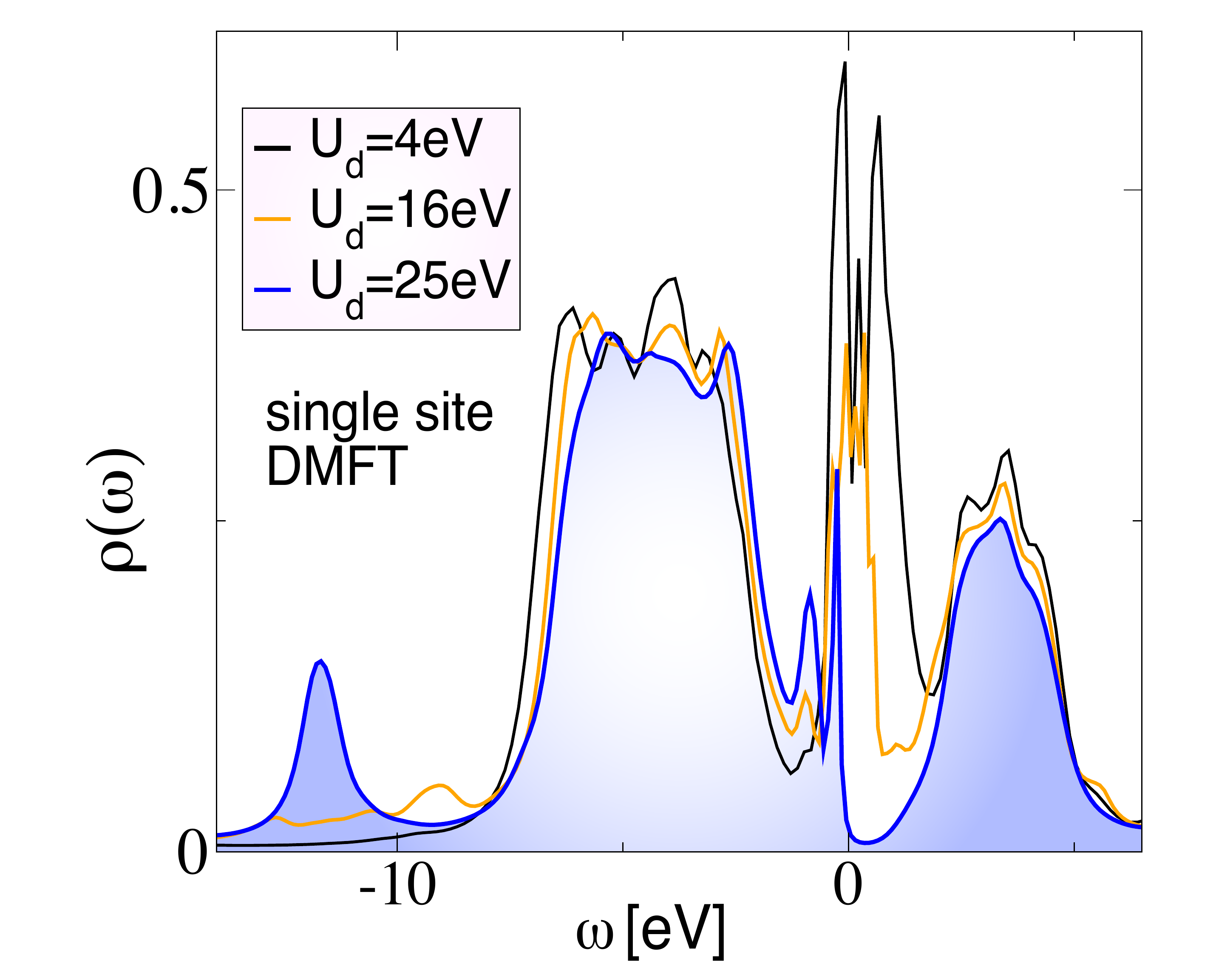}
\caption{(Color online)  
Dependence of the spectral function $\rho(\omega)$ with respect to the Coulomb repulsion $U_d$.}
\label{fig1}
\end{center}
\end{figure}
We first discuss single-site DMFT calculations 
for paramagnetic \vox. The dependence of the spectral function on the on-site repulsion $U_d$ is shown in Fig.~\ref{fig1}. 
We find that the M$_1$ phase of \vo is metallic for $U_d=4$~eV and that there is a large spectral weight present at the Fermi level. 
Hence, \vo is described by DFT+DMFT as a charge transfer correlated paramagnetic metal,
with a moderate mass renormalization $m^*/m=1.35$,
of the same order as the mass renormalization in the rutile phase obtained by other groups ($m^*/m=1.8$ 
from Ref.~\onlinecite{vo2_paper_ref_8_evidence_strong_coulomb_renormalization} and $m^*/m= 1.51$ from Ref.~\onlinecite{vo2_paper_ref11_dynamical_singlet}).
The large spectral weight at the Fermi level in Fig.~\ref{fig1} is of predominantly \dxy character, the contribution from the 
\eg orbitals being negligible: the spin-independent orbital densities at the Fermi level $ \rho_\sigma(\epsilon_F )  $ are 
0.02,0.02,0.19,0.25,0.28 for, respectively, \dxd,\dzd,\dyzd,\dxzd,\dxy symmetry, 
which indicates a strong selection of the \tg orbitals at the Fermi level in agreement with the orbital selection scenario 
argued long ago by Goodenough \cite{vo2_paper_ref_goodenough}.
Notably, we find that the dynamical correlations, described by the imaginary part of the self-energy, 
also suggest that the \dxy orbital is the most correlated orbital, whereas the \eg states are weakly correlated (for more details see Fig.~2 and Fig.~3 of the supplementary material).
We emphasize that here the oxygen $2p$ subspaces act merely as charge reservoirs, 
since the full Kohn-Sham Green's function is computed and then projected onto the correlated $3d$ subspaces.
Indeed, we find that the low energy physics is obtained by the $3d$ orbitals near 
the Fermi level, in agreement with previous observation that the spectral features in \vo clusters are reproduced by 
effective Hamiltonian with $3d$ orbitals only \cite{vo2_tanaka}. 
However, our results deviate from the description of Goodenough: 
we obtain a vanadium 3d sub-shell filling of $n=3.15$ electrons from DFT,
much larger than the 3d$^1$ configuration of the ionic picture.
We emphasize that we used a set of local Wannier orbitals, variationally optimized during the energy minimization carried out in the DFT 
calculations~\cite{onetep_ref_ngwf}, which renders the calculation of the electronic density very reliable.
Larger 3d orbital occupations in \vo than the single electron 
 have been reported in earlier DFT calculations (LSDA$+U$  
finds $n=2.48$~e \cite{vo2_paper_ref2}). We note, furthermore, 
that similar occupancies are obtained for the R phase, both by experimental 
measurement ($n=1.78$~e from XAS  \cite{vo2_paper_ref2}) and 
DFT calculations (LSDA$+U$ finds $n=2.31$~e \cite{vo2_paper_ref2} and LMTO-ASA gives $n=3.35$~e \cite{vo2_paper_ref_12_lmto_rutile_nd3}).
For larger $U_d$, we find that the spectral weight at the Fermi level shrinks, and we obtain
an insulator for $U_d=25$~eV, placing \vo well below the ZSA boundary $U_d^c$, estimated between 21~eV and 25~eV.
We conclude that a large 3d-3d Coulomb interaction alone is not sufficient to generate a large band gap for \vox,
as also suggested by early LDA calculations
\cite{vo2_paper_ref_allen_peierls_or_mott,vo2_paper_ref_10_rotation_axis} which failed to reproduce the insulating state.
Finally, we also explored the dependence on the Hund's coupling $J$ and found no significant change in the mass renormalization
by varying $J$ between $0.3$~eV and $1.2$~eV, for fixed $U_d=4$~eV, although increasing J enhances the mass renormalization for $U_d=8$~eV
and $U_d=16$~eV. 
\begin{figure}
\begin{center}
\includegraphics[width=1.0\columnwidth]{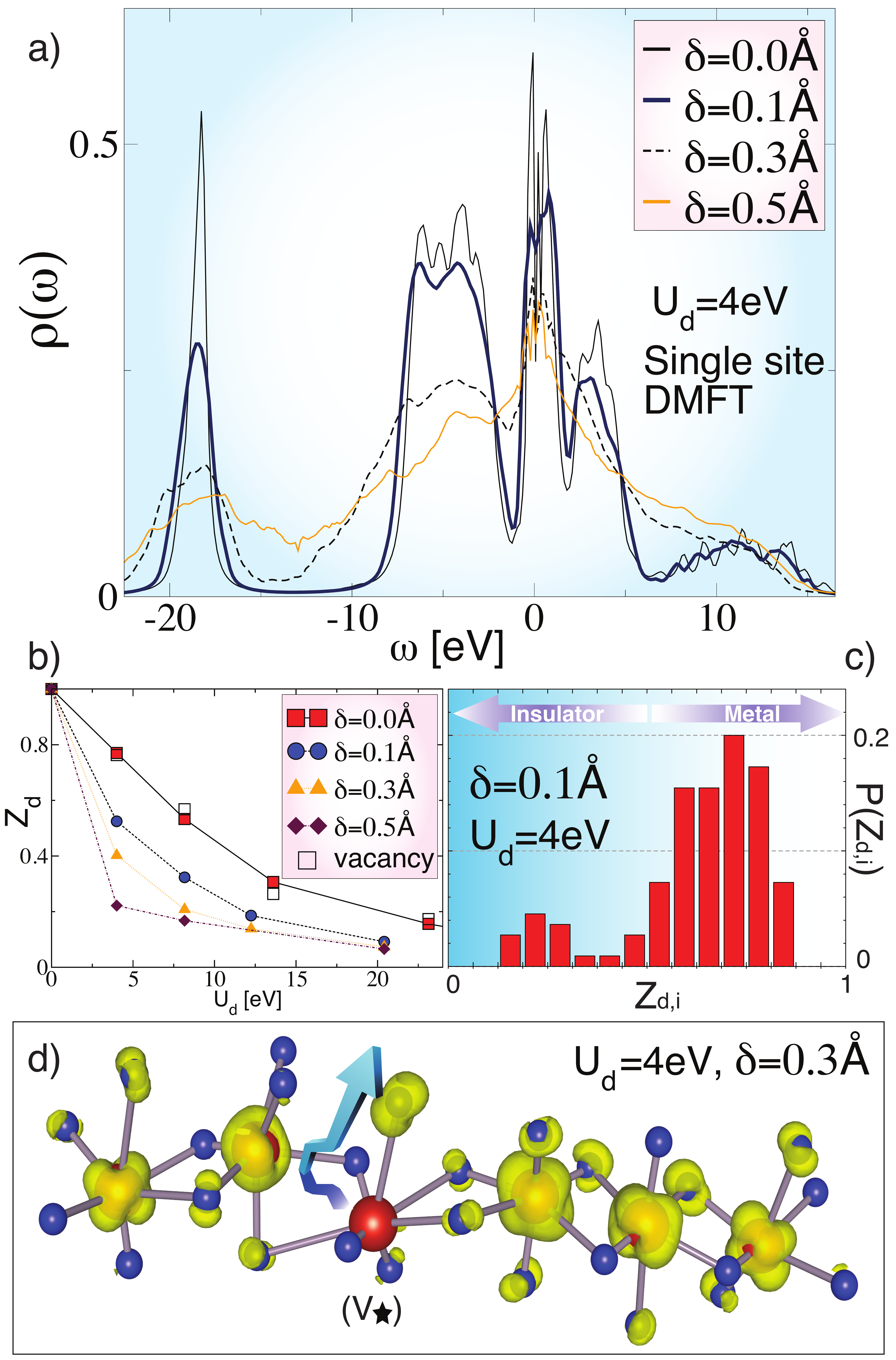}
\caption{(Color online) a) Spectral function $\rho$ of paramagnetic \vo in the presence of Gaussian disorder $\delta$. 
b) Averaged quasi-particle weight $Z_d$ with respect to the repulsion $U_d$ for
zero ($\delta=0$~\AA) to large disorder ($\delta=0.5$~\AA). Calculations including a single O vacancy in the 
$\delta=0$~\AA~case are also shown for comparison (open squares). c) 
Distribution of the local quasi-particle weight $Z_{d,i}$ for $\delta=0.1$~\AA. 
d) Isosurface of the real space representation of the Fermi density for disorder $\delta=0.3$~\AA. The large (small) sphere denotes V (O) atoms along the rutile axis.}
\label{fig2}
\end{center}
\end{figure}

If Coulomb correlations alone cannot lead to insulating behavior, perhaps 
the inevitable disorder due to imperfections of the crystal, 
or self-trapping due to strong electron-phonon coupling could be relevant. Hence we applied a random
three-dimensional Gaussian displacement to both the V and O atomic sites. 
The Gaussian width $\delta$ characterizes the amplitude of the disorder. 
The spectral function for disordered \vo is shown in Fig.~\ref{fig2}.\textbf{a}. 
Although the spectral weight at the Fermi level is suppressed with increasing disorder (reflecting charge localisation) 
the system remains metallic up to the largest physical amplitudes of disorder.
The effect of the localization induced by disorder is also observed in the
averaged quasi-particle weight $Z_d$ (Fig.~\ref{fig2}.\textbf{b}) and in the spatial distribution  
of the local quasi-particle weight $Z_{d,i}$ (Fig.~\ref{fig2}.\textbf{c}),
which clearly shows that the disorder generates regions in the crystal with strong localization, which coexist with metallic parts of the crystal
where the localization has only a weak effect. These droplets of strongly correlated Fermi liquid generate a larger mass renormalization $m^*/m$ on average,
as observed in the decrease of the averaged quasi-particle weight ($Z_d=m/m^*$) as the disorder increases (Fig.~\ref{fig2}.\textbf{b}). 
The localization effect can be understood in a simple picture: when the O atoms move closer
to the V atomic site, the static charge repulsion induces a larger charge transfer energy $\Delta=\epsilon_d-\epsilon_p$, which enhances the strength of the
correlation locally (the repulsion U of the one band Hubbard model translates into the charge transfer energy in d-p theories \cite{zsa_paper}). 
This effect is illustrated in Fig.~\ref{fig2}.\textbf{d}, where we show an isosurface of the real-space representation of the Fermi density $\rho(\epsilon_F,r)$ for one of 
the V chains along the rutile axis for $\delta=0.3$~\AA,  where large (small) spheres denote V (O) atomic sites. 
The V atom highlighted by a star has two very near
oxygen neighbors, which is expected to induce a larger charge transfer energy.  
The latter results in a transfer of charge from the vanadium site to one of its oxygen neighbors (indicated by an arrow). 
The subtle interplay between the localization induced by 
the disorder (Anderson-like) and the localization induced by strong correlations (Mott-like) is captured by the DFT+DMFT methodology.
\begin{figure}
\begin{center}
\includegraphics[width=0.9\columnwidth]{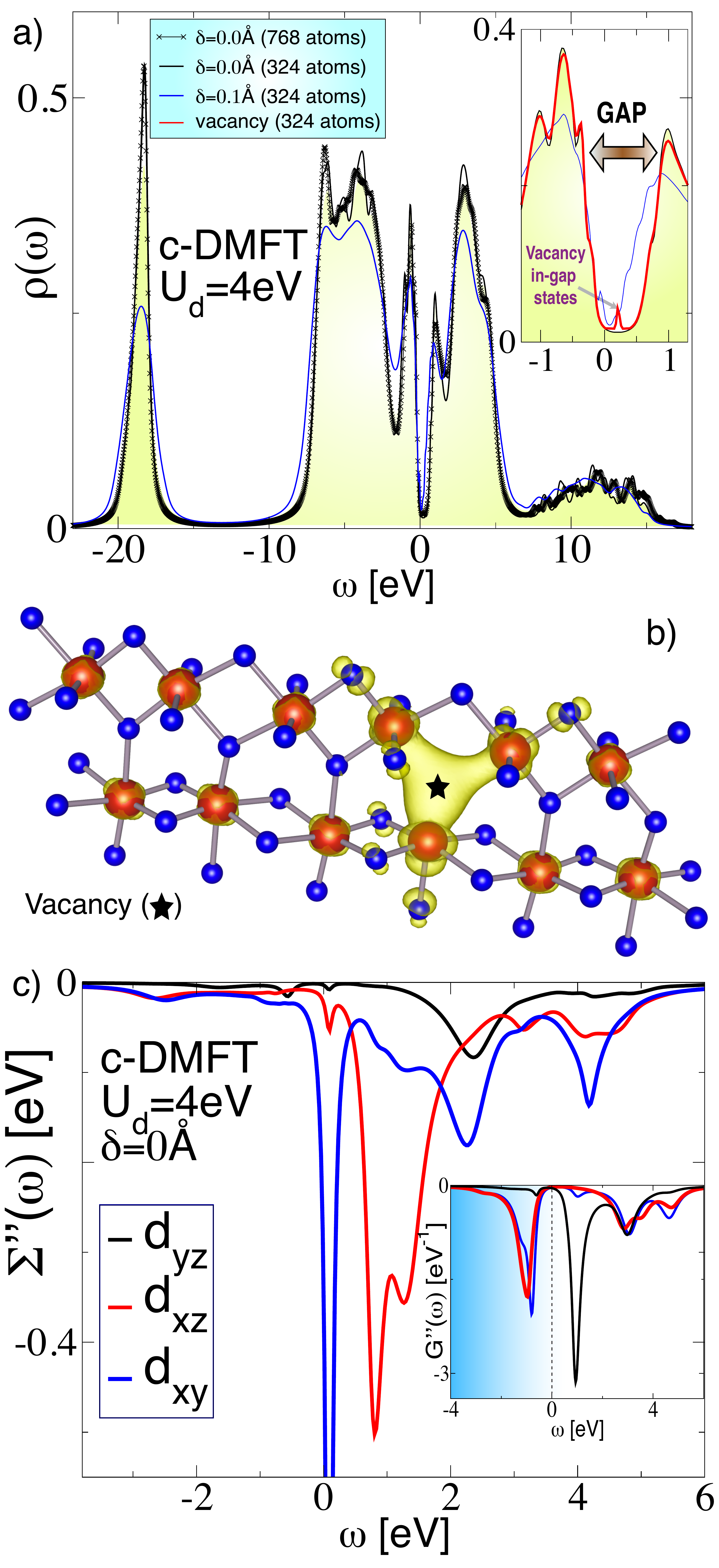}
\caption{ 
(Color online) a) Spectral function $\rho$ obtained using 
cellular cluster DMFT (c-DMFT) calculations  without ($\delta=0$~\AA) disorder
for moderate (324 atom) and large (768 atom) super-cells.
Calculations for disordered \vo  ($\delta=0.1$~\AA) and for a single O vacancy are also shown
for comparison. Inset: Enlargement of the low energy scale, the gap of $\sim 0.6$~eV is shown. The vacancy
introduces a mid-gap state, highlighted by the arrow. 
b) Isosurface of the real space representation of the charge density at the Fermi level for calculations for an O vacancy (star). 
The large (small) sphere denotes V (O) atoms along the rutile axis (horizontal direction).
c) Imaginary part of the self-energy and (inset) 
imaginary part of the Green's function for $\delta=0$~\AA.}
\label{fig3}
\end{center}
\end{figure}

We now move to the non-local cluster cellular DMFT calculations (c-DMFT). 
In c-DMFT, the cluster impurity of the AIM is mapped onto the $3d$ electron subspaces of a pair of V atoms 
forming a dimer aligned with the rutile axis. The non-locality of the self-energy between dimerized vanadium 
$3d$ subspaces is thereby self-consistently included in the calculations.
The non-local correlation present in cluster DMFT drives 
\vo to an insulator (Fig.~\ref{fig3}.\textbf{a}),
in agreement with earlier DMFT calculations using model Hamiltonians \cite{vo2_paper_ref11_dynamical_singlet}
and we obtain a gap of $\sim 0.6$~eV in agreement with both the latter and the experimental value \cite{vo2_paper_ref_gap}. 
We did not observe
any finite size effect, and the 
Peierls gaps obtained extracted 
from the 324 and 768 atom super-cells are identical. 
We find that large disorder quenches the Peierls state
for $\delta>0.1$~\AA. Very interestingly, the insulating Peierls state survives for moderate disorder $\delta=0.1$~\AA, although the gap is reduced down to $\sim 0.3$~eV.
We also carried out cluster DMFT calculations for the case of a single O vacancy: the O vacancy creates a mid-gap state 
(inset of Fig.~\ref{fig3}.\textbf{a}), spatially localized in the center
of the three V atoms surrounding the vacancy, as illustrated by the real space representation of the Fermi level
density (Fig.~\ref{fig3}.\textbf{b}) but does not strongly affect the band edges. In conclusion, our results suggest that the Peierls instability in 
\vo is very robust, surviving external perturbations such as reduction of the long-range crystallographic order or local impurities.
The imaginary part of the self-energy of the dynamical Peierls singlet is shown in Fig.~\ref{fig3}.\textbf{c}. We observe that the 
gap is mainly induced by dynamical correlations in the \dxy orbital, which exhibit a pole at the Fermi level. 
In our view, the dynamical V-V
dimers generate a Mott instability (the mechanism may be thus termed Peierls-Mott). In particular, the spectral weight (inset) shows that the cluster DMFT almost entirely depletes the 
\dyz orbital, leaving two electrons equally shared between the \dxz and \dxy orbitals. The lobe of the latter orbitals point towards the rutile axis, whereas the \dyz orbital is oriented perpendicular to this direction
and thus the latter does not contribute strongly to the orbital bonding within a V-V dimer.
Interestingly, therefore in our picture we find that two electrons per V atom lie in bonding orbitals, 
leading to a strong Mott dynamical divergence in the self-energy (Peierls-Mott). 
This contrasts with the picture of Ref.~\onlinecite{vo2_paper_ref11_dynamical_singlet}, where a single electron on each V is of bonding character and the repulsion $U_d$
drives the bonding orbitals to a singlet configuration, 
following the early proposal of Sommers and Doniach \cite{vo2_paper_ref_sommers_doniach_picture}. 
In the latter picture, the repulsion energy may be dramatically reduced by the formation of the singlet state, manifested in the fact that the low-frequency behavior of the on-site component of the self-energy, that 
associated with the \ag orbital, is linear in frequency, as opposed to a Mott insulator in which $\Sigma''$ diverges.
In our picture, the non-local self-energy affects the hybridization between intra-dimer orbitals, and act to deplete the 
\dyz orbital, leaving $2$~e in two orbitals per V site, 
in turn generating a Mott instability 
which creates a pole in the local self-energy (Peierls-Mott).  
We note that the Peierls picture in our view involves more than one electron,
due to the non-trivial hybridization between the vanadium and oxygen orbitals,
as recently obtained in XAS \cite{vo2_paper_ref2} and photoemission spectroscopy measurements \cite{vo2_paper_ref_9_photoemission}, 
which hint at an occupation of $n \approx 2$ for the 3d sub-shell in the M$_1$ phase. 
Although we find that the low energy physics is captured by the correlated $3d$ orbital subspaces alone, 
in agreement with Ref. \cite{vo2_tanaka}, we note that the oxygen $2p$ contribute indirectly to the
correlations present in the $3d$ shell by fixing the $3d$ occupation, which is captured in our fully ab inito treatment. 
\begin{figure}
\begin{center}
\includegraphics[width=1.0\columnwidth]{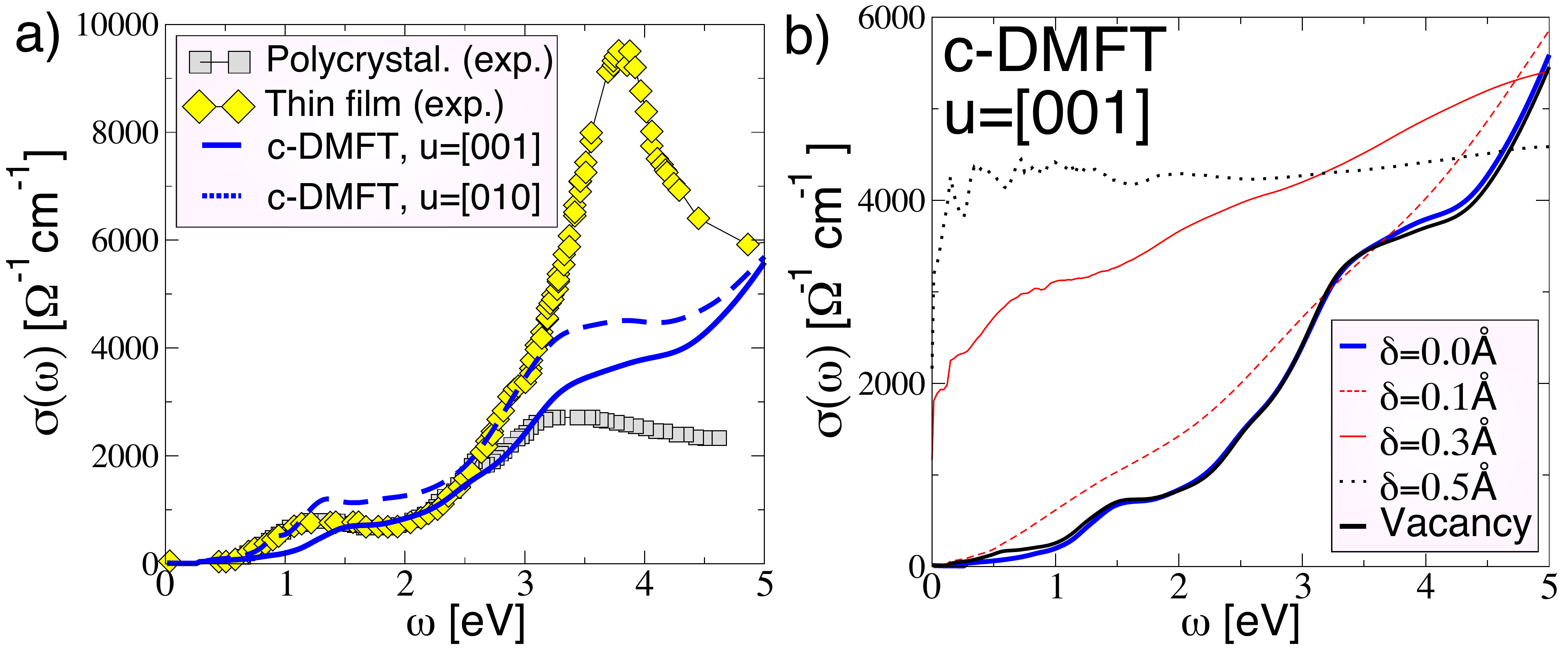}
\caption{ 
(Color online)
a) Theoretical optical conductivity along the rutile axis (solid lines) and along the perpendicular direction (dashed lines) 
calculated with cellular DMFT. Experimental data for polycrystalline films \cite{vo2_paper_ref6_optics_pollyscristaline} (squares)
and thin films (diamonds) \cite{vo2_paper_ref3} are shown for comparison.
b) Optical conductivity along the rutile axis obtained by cellular DMFT for disordered \vo for various disorder amplitudes $\delta$.}
\label{fig4}
\end{center}
\end{figure}
Finally, the optical conductivity calculated using 
cluster DMFT (Fig.~\ref{fig4}.\textbf{a}) is in 
qualitative agreement with experimental data obtained for
polycrystalline films  \cite{vo2_paper_ref6_optics_pollyscristaline} and thin films \cite{vo2_paper_ref3}.
We note that the optical gap is 
not dramatically affected by a moderate degree of
disorder $\delta=0.1$~\AA~(Fig.~\ref{fig4}.\textbf{b}). 
For large disorder, however, \vo is a bad metal and we note, in particular, that no Drude peak
is obtained in the optical conductivity, and that the disorder induces strong oscillations in the 
optical response for the infrared 
frequency range $\omega<1$~eV.
In conclusion, we have carried out linear scaling first principle calculations, in combination with cluster DMFT, on
\vox, both with and without disorder. 
We find that the ZSA boundary of the paramagnetic insulator is obtained only for unrealistic values of the Coulomb repulsion
($U_d \approx 25$~eV). 
We propose a new mechanism for the insulating M$_1$ phase of \vo based
on an orbital selective Mott transition, assisted by the Peierls distortion: 
the Peierls instability involves an orbital selection, and bonds the
\dxy and \dxz orbitals along the rutile axis, filling each orbital with one electron, and in turn
generates a Mott instability.  
This scenario may be described as \emphasize{Peierls assisted orbital selective Mott transition} 
and reconciles the simpler one electron Peierls picture with recent soft x-ray absorption spectroscopy (XAS) \cite{vo2_paper_ref2}, which points towards a breaking
of the one electron per 3d orbital picture suggested early by Goodenough \cite{vo2_paper_ref_goodenough}.
Finally, we demonstrated that the Peierls phase 
survives moderate Gaussian disorder ($\delta=0.1$~\AA),
and hence our picture accounts 
for the observation of the MIT in the experimentally realistic, disordered system \cite{disorder_vo2_no_peierls}.
Finally, we found that oxygen vacancies 
induce a localized mid-gap state, leaving the band edges 
unaffected, shedding some light on thin-film 
measurements where substrate strain can induce 
stoichiometric modification \cite{vo2_structure_power_is_better}. 
Our results, combining lattice disorder and a powerful 
method for describing non-local, dynamical correlation,
open up new frontiers for
first principles materials design under
under realistic experimental conditions.
We are grateful to Kristjan Haule for discussions and sharing his CTQMC code. 
C.W. was supported by the Swiss National Foundation for Science (SNFS).
D.D.O'R. was supported by EPSRC and the National University of Ireland.
N.D.M.H was supported by EPSRC grant number EP/G055882/1.
P.B.L is supported by the US Department of Energy under FWP 70069.


%

\end{document}


\title{Vanadium dioxide : A Peierls-Mott insulator stable against disorder.\\ Supplementary material}
\author{C\'edric Weber$^{*}$}
\affiliation{Cavendish Laboratory, University of Cambridge,
  J. J. Thomson Avenue, Cambridge CB3 0HE, United Kingdom}
\author{David D. O'Regan}
\affiliation{Cavendish Laboratory, University of Cambridge,
  J. J. Thomson Avenue, Cambridge CB3 0HE, United Kingdom}
\affiliation{Theory and Simulation of Materials, 
\'{E}cole Polytechnique F\'{e}d\'{e}rale de Lausanne,\\ 
Station 12, 1015 Lausanne, Switzerland}
\author{Nicholas D. M. Hine}
\affiliation{Cavendish Laboratory, University of Cambridge,
  J. J. Thomson Avenue, Cambridge CB3 0HE, United Kingdom}
\affiliation{The Thomas Young Centre and the
Departments of Materials and Physics, 
Imperial College London, 
London SW7 2AZ, United Kingdom}
\author{Mike C. Payne}
\affiliation{Cavendish Laboratory, University of Cambridge,
  J. J. Thomson Avenue, Cambridge CB3 0HE, United Kingdom}
\author{Gabriel Kotliar}
\affiliation{Rutgers University, 136 Frelinghuysen Road, Piscataway, NJ, U.S.A.}
\author{Peter B. Littlewood}
\affiliation{Cavendish Laboratory, University of Cambridge,
  J. J. Thomson Avenue, Cambridge CB3 0HE, United Kingdom}
\affiliation{Physical Sciences and Engineering, Argonne National Laboratory, Argonne, Illinois 60439, U.S.A.}


\maketitle
 
   
\section{Methodology}

Kohn-Sham density-functional theory (DFT)~\cite{dft_kohn, kohn_sham} has been shown to be capable of making accurate predictions for many materials.
DFT combines high accuracy and moderate computational cost, yet the computational effort required to perform
calculations with conventional DFT approaches is still non negligible: it increases with the cube of the number of atoms,
rendering them unable to routinely tackle problems comprising more than a few hundred atoms, even on modern large supercomputers.
Since the pioneering work of the Nobel laureate Walter Kohn~\cite{lin_scaling_dft}, 
it has been known that it is possible to 
reformulate DFT to exploit the short-ranged
nature of quantum mechanics, so that the computational cost scales linearly which would, in 
principle, allow calculations with many thousands of atoms.
The ONETEP approach~\cite{onetep,linearscalingdftu,onetep_ref6}, used in this 
work, is an example of the practical 
realization of linear-scaling DFT and DFT+$U$, and is particularly
notable for its accuracy equivalent to a plane-wave method,
by means of its \emph{in situ} optimization of a set of local 
Wannier  orbitals with a respect to a systematically improvable basis.

The linear scaling performance of ONETEP allowed us to study large atomic super-cells. 
Specifically, we investigated both a moderate
supercell of 324 atoms for \vo (108 V atoms) and a large supercell of 768 atoms (256 V atoms). 
We used, for the crystallographic structure of the M$_1$ phase of \vox, the lattice constants and atomic positions obtained for powder \vo ($a=5.743$~\AA, $b=4.517$~\AA, 
$c=5.375$~\AA, and $\beta=122.6^\circ$)
since the lattice parameters of thin films vary with charge localization and substrate-induced built-in strain~\cite{vo2_structure_power_is_better}.
The electronic core states were represented by 
GGA scalar-relativistic pseudo-potentials generated with the OPIUM package~\cite{opium_package}. 
The (3s,3p,3d,4s) and (2s,2p) atomic shells were respectively treated as valence states for the V and O atoms.
Hence, we used 10 
orbitals to describe each V atom and 4 to describe each O atom,
giving a total of 1944 orbitals for the 324 atom super-cell and 4608 for the 768 atom super-cell.

ONETEP~\cite{onetep} is at the cutting 
edge of developments in first principles calculations.
However, while the fundamental difficulties of performing
accurate first-principles calculations with 
linear-scaling cost have been solved,
one remaining problem of DFT approaches in general is that conventional approximations
to the exchange-correlation (XC) functions fail in describing many compounds where
strong correlations are present. In such cases, DFT 
often predicts moments and energetics that are qualitatively inconsistent with experiment, or fails to describe
the insulating state and converges to a metallic state instead.
DFT+DMFT is a computationally moderately expensive method for refining 
the description of on-site Coulomb interactions provided by such XC functionals and, hence, for extending the range of applicability of DFT to strongly-correlated materials.

The principle of DFT+DMFT is to separate the dominant subspace
of the system, which is well-described by conventional XC functionals due to its delocalized and free-electron character,
from a subset of localized subspaces, so-called \emphasize{correlated sites}, which are strongly correlated and not adequately 
described by the XC functionals. The XC correlated sites are then explicitly augmented with screened Coulomb and Hund's coupling interactions within these sites, 
together with a double-counting term to correct for the component already included within the DFT XC functional.

The correlated sites are typically spanned by a set of 
3d and/or 4f atomic-like orbitals, which define a set of localized Hubbard projectors \cite{PhysRevB.82.081102}.
Solutions of appropriate orbital symmetry of the hydrogenic Schr\"odinger equation, 
such as atomic-like or linear muffin-tin orbitals, are
the most common choice~\cite{vo2_paper_projector1,vo2_paper_projector2,vo2_paper_projector_3} for Hubbard projectors.
The vanadium-centered correlated sites were delineated in our calculations using hydrogenic 3d orbitals 
as Hubbard projectors, characterized by the canonical Clementi-Raimondi~\cite{clementi_raimondi} effective charge of $Z=8.9829$
for this atomic sub-shell and species, determining their spatial diffuseness. 

In the implementation of the linear-scaling DFT method used in this work, 
we work with the single-particle density-matrix, 
which is expressed in a separable form~\cite{vo2_paper_ref_advance_in_dft_separable_form, linear_scaling_separable_form_density_onetep}
\footnote{We use the Einstein convention to simplify the notations, in particular repeated Greek indices
implicitly sum over the NGWFs, and as per the Einstein convention, parenthesis suspend the summation implied by
matching pairs of indices.}:
\begin{equation}
\rho \left( \mathbf{r}, \mathbf{r'} \right)
=  \sum\limits_{\alpha \beta} 
{\phi_\alpha \left(  \mathbf{r} \right) K^{\alpha \beta} \phi_\beta
\left(  \mathbf{r'} \right)},
\end{equation}
where $ \left\lbrace \phi_\alpha \right\rbrace$ 
is a set of localized nonorthogonal generalized Wannier functions (NGWFs) variationally optimized during the energy minimization carried out as part of the DFT calculations~\cite{onetep_ref_ngwf}.
The	density kernel $K^{\alpha \beta} = \langle  \phi^\alpha | \hat{\rho} |\phi^\beta \rangle $ is the 
representation of the single-particle density operator $\hat \rho$ in terms 
of the contravariant NGWF duals $\left\lbrace \phi^\alpha
\right\rbrace $, defined as those which satisfy the
relationship $\langle \phi^\alpha | 
\phi_\beta \rangle= \delta_{\alpha \beta }$.
The NGWFs are, in turn, expanded in terms of a systematic basis of Fourier-Lagrange, or psinc functions~\cite{onetep_optimisation_ngwfs}. 
The size of this basis is determined by a plane-wave kinetic energy cutoff (a minimum of 822~eV for all reciprocal lattice vectors in our calculations). 
The DFT energy functional is iteratively minimized with respect to both the density kernel and the NGWF expansion coefficients using a minimization scheme 
which is detailed in Ref. \onlinecite{onetep_energy_min}.
Therefore, the NGWFs are a readily accessible set of localized orbitals 
(spatially truncated to atom-centered spheres of radius $6.6$~\AA)
which are calculated with linear-scaling computational cost.
No additional approximation such as the density kernel truncation was used in this work, and the energy convergence was tested against
the plane-wave kinetic energy cutoff and the spatial truncation of the NGWFs.

Once the fully converged energy minimization of \vo was carried out, the full Green's function was computed in the finite temperature Matsubara 
representation~\footnote{A fixed temperature T=189 K was used throughout all our calculations} 
from the readily available DFT Hamiltonian $\bold{H}$:
\begin{equation}
\label{fullg}
G^{\alpha \beta} \left(i \omega_n \right) = \left( ( i \omega_n+\mu )S_{\alpha \beta}  - H_{\alpha \beta} -  \Sigma_{\alpha \beta}  \right)^{-1},
\end{equation} 
where $\mu$ is the chemical potential (fixed at the mid-energy between the last occupied state of the ONETEP calculation and the first empty state),
$\bold{S}$ is the full overlap matrix between the NGWFs so that
$\phi_\alpha = S_{\alpha \beta} \phi^\beta$ 
and $\bold{\Sigma}$ is the self-energy matrix computed 
by the DMFT algorithm. We note that the computation of the Green's function requires large matrix inversion (in particular, for \vox,
the matrix to invert is of size $1944\times1944$ for the 324 atom super-cell and $4608\times4608$ for the 768 atom super-cell. 

\begin{figure}
\begin{center}
\includegraphics[width=0.9\columnwidth]{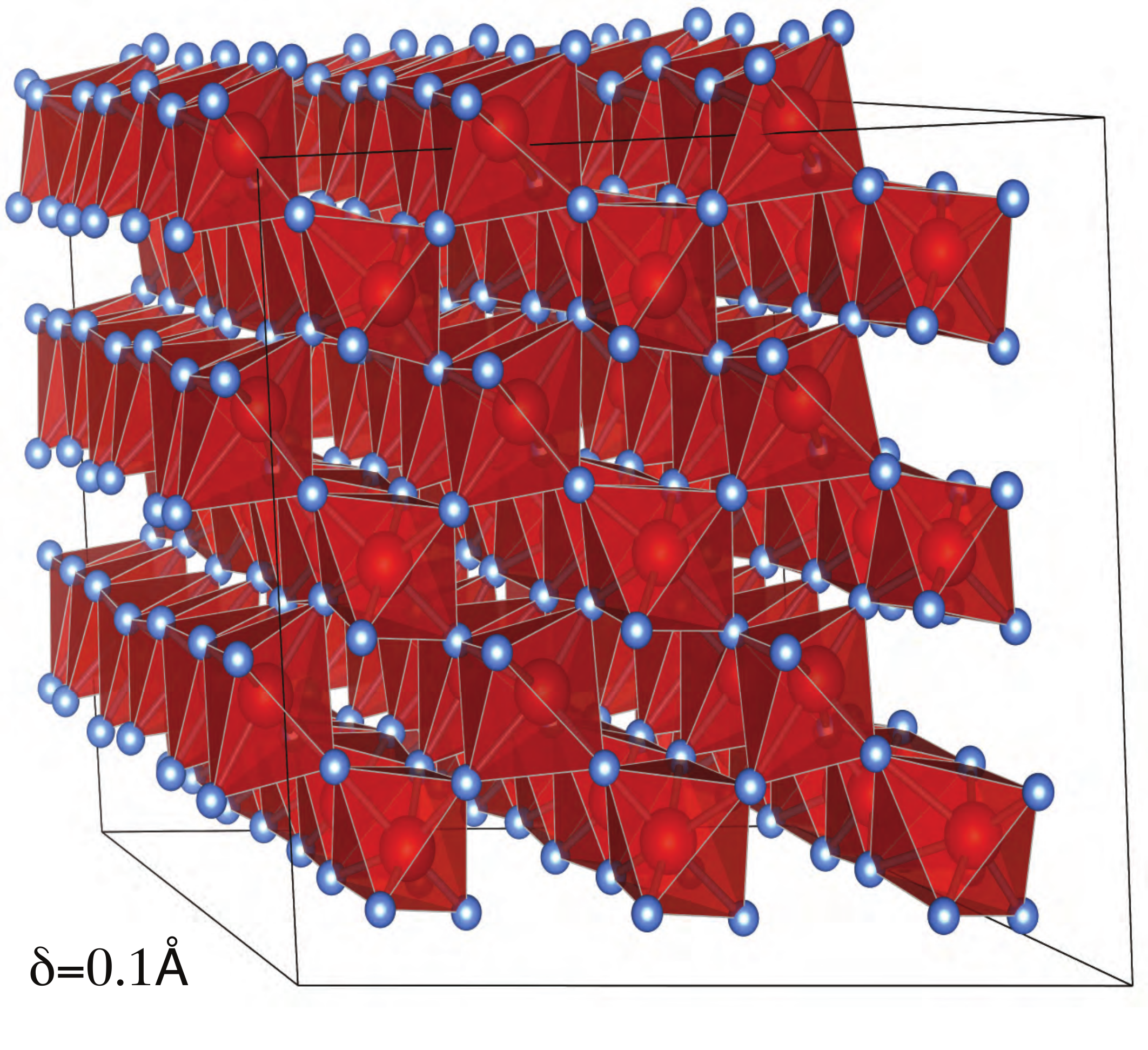}
\caption{
{\bf Super-cell for DFT+DMFT calculations: } (Color online) 
Super-cell used for the 324 atom DFT+DMFT calculations 
for disordered VO$_2$ with $\delta=0.1$~\AA.}
\label{fig5}
\end{center}
\end{figure}

These super-cells allowed us to describe spatially dependent perturbations, 
such as static disorder (see Fig.~\ref{fig5}), and also to consider oxygen vacancies \footnote{We note that we performed the vacancy calculations with a neutral simulation cell, i.e. we did not add or remove any electrons with respect to the valence given by the norm-conserving pseudo-potentials.}.
In particular, we applied in our calculations a random
three-dimensional Gaussian displacement to both the V and O atomic sites. 
The Gaussian width $\delta$ characterizes the amplitude of the disorder. 
We note that a large number of uncontrolled and
uncorrelated experimental parameters, such as the variation of the
atomic positions due to the shear induced by dislocations, are
described at the theoretical level by a Gaussian distribution (central
limit theorem). Notwithstanding, we use a single representation of disorder for each amplitude $\delta$, 
Gaussian disorder satisfying the ergodicity theorem such that the local average over different configurations is equivalent to the average 
over space for large enough super-cells.

Matrix inversion was carried out
for $100$ Matsubara frequencies to provide converged sampling of the Green's function.
To achieve reasonable computational time, we performed the matrix inversion
on recently-developed graphical computational units (GPUs) using a home-made parallel implementation of the Cholesky LU decomposition with the CUDA programming language. 
The calculations were carried out using 8 GPUs in parallel, achieving a total of 4000 GFLOPS (giga floating-point operations per second), allowing a tremendous speed-up of Green's function calculation (the computational time needed to carry out the matrix products and matrix inversions involved in the calculation for a 
100 Matsubara frequencies was of the order
of a few minutes for the largest super-cell). 
The matrix multiplications involved throughout the 
DMFT calculations were also performed using the CUDA architecture.

The projected Green's 
function $\bold{\tilde{G}}$ of the 
vanadium 3d correlated sites is obtained as
\begin{equation}
\label{projgreen}
{\tilde G}^ {\left(I\right)}_{  m m'} \left(i \omega_n \right) = W_{ m \alpha}^{\left( I \right)  }  G^{\alpha \beta}  \left(i \omega_n \right) V_{ \beta m'}^{\left( I \right) },
\end{equation}
where $I$ runs over the Vanadium atoms,
$m$ and $m'$ over the five atomic 3d orbitals used as orthonormal Hubbard projectors 
(in real cubic-harmonic notation: \dxd, \dzd, \dyzd, \dxzd, \dxyd), $\alpha$ and $\beta$ are the indices for the NGWFs,
and the matrices ${V}^{\left(I\right) }_{\alpha m} = 
\langle  \phi_\alpha | \varphi^{\left( I \right)}_m \rangle$ and ${W}^{\left(I\right) }_{m \alpha} = 
\langle  \varphi^{\left( I \right)}_m |  \phi_\alpha \rangle$ are the overlap between the
NGWFs $\left\lbrace \phi_\alpha  \right\rbrace$ and the 
hydrogenic orbital $\{ \varphi^{\left( I \right)}_m \}$ at the Vanadium site $I$ and with orbital index $m$.
Note that the orthonormality of the Hubbard projectors implies that we do not need to 
distinguish them from their duals (the generalization to nonorthogonal projectors is discussed in Ref~\onlinecite{dftu_david_subspace_representation}).

The strong correlation acting within each Vanadium site is described by the conventional
Slater-Kanamori form of the interaction term~\cite{slater_kanamori_interaction,hund_coupling_kanamori},
namely 
\begin{align}
\label{hint}
\mathcal{H}_{int} ={}& U_d \sum \limits_{m} {n_{m \up} n_{m \dn}} + (U'_d-\frac{J}{2})\sum\limits_{m>m'}{n_{m}n_{m'}}  \\
\nonumber {}&
-J \sum\limits_{m>m'}{\left( 2 {S}_m {S}_{m'} + \left(  d^\dagger_{m\up} d^\dagger_{m \dn } d_{m' \up} d_{m' \dn}   \right) \right)    },
\end{align}
for each correlated site and with no inter-site interactions, 
where the first term describes the effect of 
intra-orbital Coulomb repulsion $U_d$ and the second
term describes the inter-orbital repulsion $U'_d$. The latter is renormalized by J in order
to ensure rotational invariance of the interaction (for a review see Ref.~\onlinecite{imada_mott_review}). 
The third term is the Hund's rule coupling, described by a spin exchange of amplitude $J$.
$\mathbf{S}_{m}$ denotes the spin of orbital $m$,
$S_{m}=\frac{1}{2}d_{m s}^{\dagger}
\vec{\sigma}_{s s'}
d_{m s' }$, where $\vec{\sigma}$ is the vector of Pauli matrices
indexed by $s$ and $s'$.
Generally,	 the hierarchy of interactions is $U_d > U'_d > J$.
$U_d$ and $J$ were set to typical values for the screened interactions for \vox, 
namely $U_d=4$eV and $J=0.68$eV  \cite{vo2_paper_U,vo2_paper_J}.

The interaction term (\ref{hint}), combined with 
(\ref{fullg}) and (\ref{projgreen}), defines a local problem at each Vanadium site $I$ of the lattice,
which we solve with the DMFT algorithm~\cite{OLD_GABIS_REVIEW}. 
In the single site DMFT approach, the projected self-energy matrix $\bold{\tilde \Sigma}^{\left( I \right)
\left( J \right)}$  is assumed to be local, i.e.,
\begin{equation}
\bold{\tilde \Sigma}^{\left( I \right) \left( J \right)}=0
\end{equation}
for non-local Vanadium site indices $\left( I \neq J \right)$, unlike the cluster cellular DMFT \cite{cdmft_reference_cluster_dmft} (c-DMFT) approach 
where spatial non-local contributions from vanadium sites belonging to the same dimer are included.

The self-energy is obtained by solving an auxiliary Anderson impurity problem (AIM) for each V atom (V-V dimer) in the single DMFT (cluster DMFT). 
The hybridization matrix $\bold{\Delta}$ in the orbital subspace (\dxd, \dzd, \dyzd, \dxzd, \dxyd) of the AIM problem is defined, for each site, as:
\begin{equation}
\label{hybrid}
\bold{\Delta}(i\omega_n) = \left( i \omega_n+ \mu \right) 
\bold{\tilde O}  - \bold{\tilde \Sigma} - \bold{E^{imp}} - \bold{\tilde G}^{-1},
\end{equation}
$\bold{\tilde O}$ is the local projected overlap matrix, which is related to the non-orthogonality of the DFT basis set (NGWFs)
\cite{non_orthogonal_dmft,non_orthogonal_dmft_kotliar,non_orthogonal_basis_cdmft}
and $\bold{\tilde O} = \bold{1}$ in the particular case of orthonormal Hubbard projectors in the form of Kohn-Sham Wannier functions.
In particular, we find that $\bold{\tilde O}$ is related to the NGWF's overlap matrix by
\begin{equation}
{\tilde O}^{\left( I \right)}_{m m'} =
\left( 
W_{m \alpha}^{\left( I \right) } 
\left( S^{-1} \right)^{\alpha \beta} 
V_{ \beta m'}^{\left( I \right) } 
\right)^{-1}.
\end{equation}
The static part of the hybridisation function $\bold{E}^{imp}=\bold{\Delta} ( i \omega_n \rightarrow \infty)$ defines the energy levels of the 
impurity in the absence of hybridisation (atomic limit).
Both $ \bold{\tilde O} $ and $\bold{E^{imp}}$ 
define the impurity problem and they are
constructed such that the hybridization matrix $\bold{\Delta}(i\omega_n)$ 
exhibits the correct physical decay $ \propto 1/ i\omega_n$ at large frequency
(we ensured that $\bold{\tilde O} ^{\left( I \right) }_{m m'} = 
\lim_{  \omega \rightarrow \infty} 
\left[  \bold{\tilde G}^{-1} \left( i \omega \right)'' /\omega \right] $
to a high precision, as required by this condition).
The self-energy 
$\bold{\tilde \Sigma}$ 
is obtained by solving the Anderson impurity problem defined by the hybridization (\ref{hybrid}) 
and the interaction Hamiltonian (\ref{hint}). 

The AIM is 
solved using two different solvers: 
(1) a finite temperature Lanczos algorithm~\cite{medici_finite_temperature_lanczos,finite_temperature_lanczos_leibsch,our_plaquette_paper},
 and some of the results were cross-checked and benchmarked with (2) a continuous-time Monte Carlo
solver (CTQMC)~\cite{werner_ctqmc_algorithm,Haule_long_paper_CTQMC}. 
The two solvers both suffer from approximations but complement each other: the Lanczos solver uses a finite discretization of the hybridization (\ref{hybrid}) and suffers from finite size effects. The CTQMC
solver suffers from the so-called \emphasize{fermionic quantum sign problem} when the off-diagonal elements are considered, hence in this work we neglect within the CTQMC implementation 
the off-diagonal elements of the hybridization matrix (\ref{hybrid}).
We note, however, that the amplitude of the off-diagonal elements depends strongly 
on the choice of the localized Hubbard projectors, and, in particular, they depend on the choice for the local orientation of the local (e$_x$,e$_y$,e$_z$) axis.
The off-diagonal elements of the hybridization and Green's function 
prove to be small for the vanadium 3d set employed in this work when the local coordinates defined for \vo are used (see discussion hereafter).

Once the local self-energy 
$\bold{\tilde \Sigma}$ 
is obtained, it is up-folded to the space spanned by the
Kohn-Sham orbitals
(equivalently the space spanned by NGWFs),
in the separable form
 \begin{equation}
  \Sigma_{\alpha \beta} =
  \sum\limits_{I,J} 
  V_{ \alpha m}^{\left( I \right) }
\tilde{\Sigma}^{\left( I \right) \left( J \right) }_{m m'} 
 W_{ m' \beta}^{\left( J \right)} .
 \end{equation}
 As pointed out recently in \onlinecite{dmft_lda_implementation}, the separable form of the projectors of the self-energy enforces the causality of the
up-folded self-energy 
(the local self-energy in the 
Hubbard projector basis 
is causal by construction, so that ${\tilde \Sigma_{mm}^{(I)(I)}(i \omega_n)}'' \le 0$, $ \forall \left( i\omega_n, m, I \right) $,
since it is obtained by solving the AIM in the local orthogonal hydrogenic basis).
We note that the causality is a necessary requirement to 
obtain physical observables.
We carefully checked that all of
the obtained self-energies were causal, 
since numerical error might still induce
some unexpected fluctuation in 
the imaginary part of the self-energy). 

One particular difficulty arises when 
one wants to construct a non-interacting Hamiltonian 
${\bold \mathcal{H}}$ from the DFT-LDA ground-state density. 
Since the LDA already contains the influence of the Coulomb interaction to a certain degree, the problem of double counting of these contributions by 
the interaction Hamiltonian (\ref{hint}) arises. In order to avoid this double counting, one has to subtract off the interaction contributions from the LDA. 
Unfortunately, the precise form 
for a particular set of orbitals is not known and the best one can do is to account for these contributions in an averaged way. 
In this work, we used the canonical 
form of the correction $E_{\mathrm{dc}}$, 
given by
\begin{equation}
\label{dc}
E_{\mathrm{dc}}= U_d^{av} \left( n_d - 0.5 \right) - 
\frac{J}{2} \left( n_d - 1 \right),
\end{equation}
where $n_d$ is the total occupancy 
of the vanadium 3d subspace, as defined by the
Hubbard projectors, and  $U_d^{av}$ 
is the averaged repulsion related 
to the intra-orbital and inter-orbital repulsion, 
computed as follows~\cite{hund_coupling_averaged_double_counting}:
\begin{equation}
U_d^{av}= \frac{U_d+ 2 \left( N_{\mathrm{deg}} -1 \right) U_d'}{2N_{\mathrm{deg}}-1}
\end{equation}
and $N_{\mathrm{deg}}$ is given by the number of orbitals.

The upfolded self-energy, in the covariant 
NGWFs representation, is hence corrected by $E_{\mathrm{dc}}$,
giving
\begin{equation}
\label{sigmadc}
  {\Sigma}_{\alpha \beta} =
  \sum\limits_{I J} 
  V_{ \alpha m}^{\left( I \right)  }
\left( \tilde{\Sigma}^{\left( I \right) \left( J \right) }_{m m'} 
- E_{\mathrm{dc}}  \delta^{\left( I \right) \left( J \right)}_{m m'} 
\right)
 W_{m' \beta}^{\left( J \right)  } .
 \end{equation}
 The double-counting correction avoids spurious artifacts, such as a shift of the subspace occupancies 
when the Coulomb interaction $U_d$ is increased from zero. 
In this work, we did not find any significant deviation from the DFT density in the DFT+DMFT scheme, providing a strong
indication that the double-counting term (\ref{dc})
correctly accounts for interactions already included
in the DFT-LDA Hamiltonian $\bold{H}$.

For the particular case of \vo in the M$_1$ phase, due to the buckling of the V atoms, it is convenient to carry out the DMFT in a rotated basis, making use of 
a local coordinate system associated with the vanadium sites. 
In particular, the rotation used in this work 
is such that the local $e_x$ axis is directed towards
the rutile axis and the local $e_y$ axis points towards the in-plane oxygen. The vanadium-apical 
oxygen vector forms a 90$^\circ$ angle
with the rutile axis~\cite{vo2_paper_ref_10_rotation_axis}. This allows us to obtain an hybridization 
matrix (\ref{hybrid}), which is as diagonal as possible. 
This is crucial for the CTQMC algorithm, since it neglects the off-diagonal terms. The rotation is carried out in orbital space~\cite{rotation_harmonics} and translates  for the projected Green's function as follows:  
\begin{equation}
\bold{\tilde G^{rot}}=   \bold{\tilde U}^\dagger  \bold{\tilde G} \bold{\tilde U} ,
\end{equation}
where $\bold{\tilde U}$ is the $5\times5$ rotation matrix in cubic harmonic space, corresponding to a real space rotation $R^{(3)}$, and 
$\bold{\tilde G^{rot}}$ is the rotated Green's function. The same holds for the projected self-energy obtained by solving the AIM model  $\bold{\tilde \Sigma^{rot}}$ , which is rotated back to
the original system of coordinates:
\begin{equation}
\bold{\tilde \Sigma}=  \bold{\tilde U} \bold{\tilde \Sigma^{rot}} 
\bold{\tilde U}^\dagger.
\end{equation}
Equations (\ref{fullg}), (\ref{projgreen}),(\ref{hybrid}) and (\ref{sigmadc}) are iterated until DMFT convergence is obtained.
Real frequency quantities are readily available from the Lanczos solver and provides us access to the physical observables of the system. In particular, the density of states is available
from the retarded full Green's function:
 \begin{equation}
 \rho \left( \omega \right) =    \rho^{\alpha \beta} \left(\omega \right)  S_{\beta \alpha} ,
 \end{equation} 
 where the NGWF-resolved spectral density $ \rho_{\alpha \beta} \left(\omega \right)$ is defined by the retarded Green's function obtained at the DMFT self-consistency $\bold{G_R}$ as: 
 \begin{equation}
  \label{density}
 \rho^{\alpha \beta} \left(\omega \right)
 =\frac{  G_R^{  \alpha \beta} \left(\omega \right)
 - G_R^{ \dagger \alpha \beta} \left(\omega\right) 
  }{ 2i\pi }.
 \end{equation}
 
A very common experimental probe is optical spectroscopy, which is accessed in this work through the optical conductivity.
The optical conductivity in DFT+DMFT is computed by using the Kubo formula of linear response theory
in the \emphasize{no-vertex-corrections} \footnote{Vertex-corrections vanish for a single orbital in infinite dimensions, but for all other cases this is 
an approximation.} approximation~\cite{millis_optical_conductivity_review}, via the expression
\begin{align}
 \sigma(\omega) ={}&  \frac{2 \pi e^2 \hbar }{\Omega }  \int  
 d \omega' \;
 \frac{  f( \omega'-\omega)-f \left( \omega' \right)}{\omega} \\
 {}&
 \times 
 \left( \bold{\rho^{\alpha \beta}} \left( \omega' - \omega \right)  \bold{v}_{\beta \gamma}  \bold{\rho^{\gamma \delta}} \left(\omega' \right) \bold{v}_{\delta \alpha} \right), \nonumber
\end{align}
where the factor two accounts for the spin degeneracy, $\Omega$ is the simulation-cell volume, $e$ the electron charge,
$ \hbar $ the reduced Planck constant,
$f \left( \omega \right)$ is the Fermi-Dirac distribution,
and the optical conductivity is obtained by the convolution of charge
density matrix operator $\bold{\rho}^{\alpha \beta}$, defined in equation (\ref{density}),
and the bare vertex $\bold{v}_{\alpha \beta}$ is given for a non-orthogonal basis by~\cite{optical_conductivity_non_orthogonal_basis}
\begin{equation}
\bold{v}_{\alpha \beta}= \bold{p} / m_e  = -i \hbar / m_e 
\langle \phi_\alpha \rvert  \bold{\nabla} \lvert \phi_\beta \rangle.
\end{equation}
We emphasize that 
the so-called \emphasize{Peierls substitution} \cite{millis_optical_conductivity_review} approximation was not used here. 

Another interesting physical quantity probed in this work is the quasi-particle weight of a local vanadium 3d subspace 
$Z_{d,i}^m$, where the index $i$ runs over the Vanadium site and $m$ over the d orbitals. 
We used the discrete Eliashberg estimate to obtain the quasi-particle weight: 
\begin{equation}
Z_{d,i}^m =
\left( 1 - \frac{
\tilde{\Sigma}''^{\left( i \right)\left( i \right)}_{m m} 
\left(i \omega_1 \right)
}{\omega_1 } \right)^{-1},
\end{equation}
where $\omega_1$ is the first Matsubara frequency.
The quasi-particle weight $Z_{d,i}$ of the Vanadium site $i$ is
estimated by the Fermi liquid specific heat of each independent d orbital,
that is~\cite{fermi_velocity_LSCO_renormalization_bare_mass}:
\begin{equation}
\gamma_{i}^m \propto  \frac{\rho_i^m(\epsilon_{F}) }{ Z_{d,i}^m },
\end{equation}
where $\epsilon_F$ is the Fermi energy, and it follows:
\begin{equation}
Z_{d,i} =  \sum\limits_m{\rho_i^m(\epsilon_F)} /  \sum\limits_m{\gamma_{i}^m}. 
\end{equation}
The effect of the localization induced by disorder can be quantized by the average $Z_d$, which 
is obtained along the same lines:
\begin{equation}
Z_d =  \sum\limits_i {\rho_i(\epsilon_F)} /  \sum\limits_i{\gamma_{i}}. 
\end{equation}
$Z_d$ is related to the mass renormalization ($Z_d=m/m^*$).

Finally, we note that the DFT+c-DMFT implementation is formally very similar to the early work of P. Fulde and collaborators \cite{fulde_cluster,fulde_abinitio1}. 
In the latter, the electron-nucleus and electron-electron interactions is constructed in a fully ab initio manner, and the non-local exchange interaction 
is calculated over a finite range by projecting the full environment to a local cluster. The orthogonality of the orbitals of the short-range 
environment to those of the central cluster is achieved by including projection operators in the Fock operator, 
which also has the effect of localizing these orbitals. The latter method does provide both ground state properties and 
was extended to finite temperature excitations  \cite{fulde_abinitio1}, and gives a fully ab initio perspective.


\section{Comparison between the ED and CTQMC impurity solvers}

\begin{figure}
\begin{center}
\includegraphics[width=1.0\columnwidth]{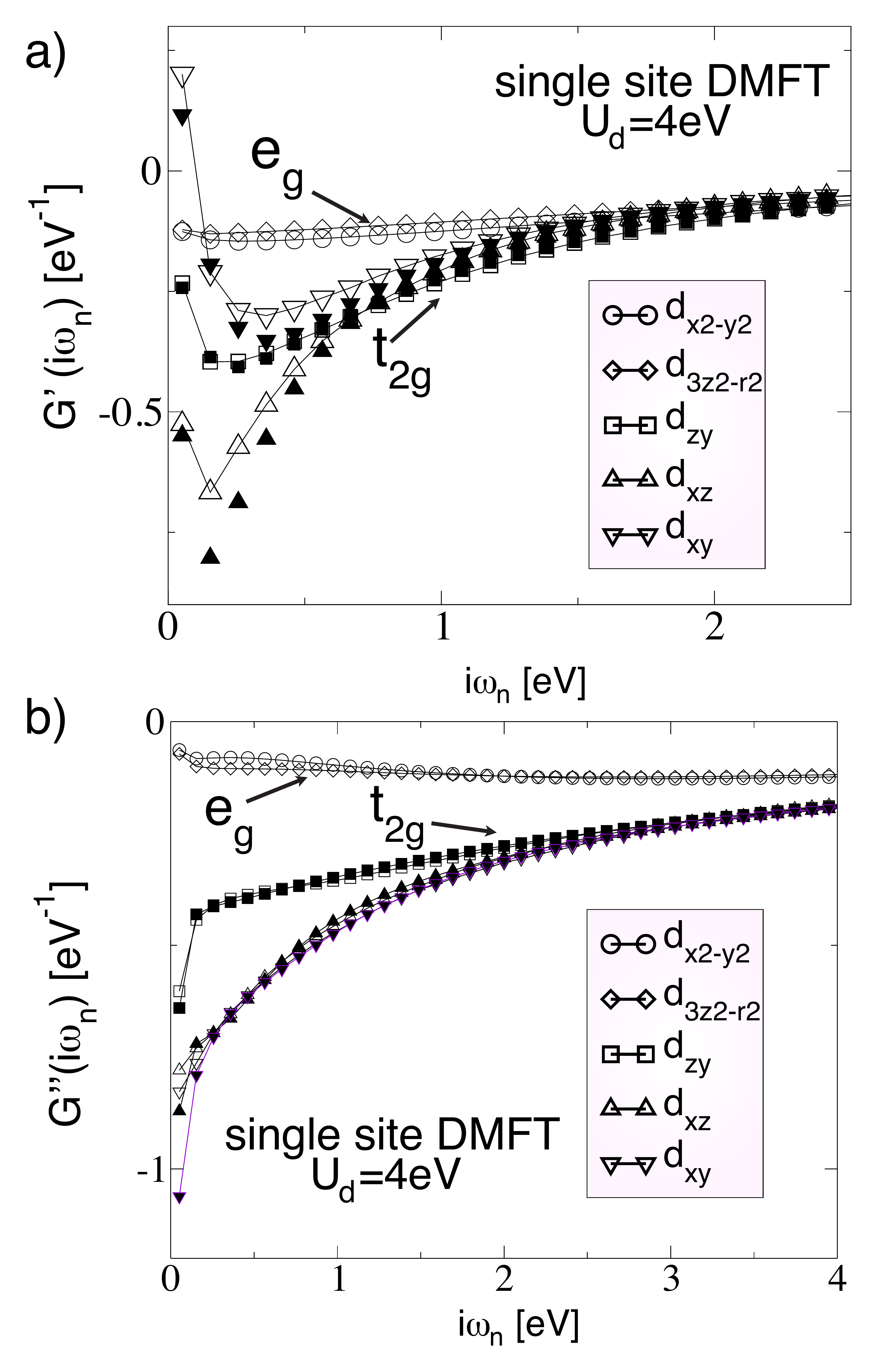}
\caption{ 
{\bf Matsubara representation of the Green's function:} (Color online) Real (a) and imaginary part (b) of the Green's function
$\bold{ \tilde{G}}$
in the Matsubara representation for paramagnetic \vox. The calculations were performed using the CTQMC impurity solver (open symbols) including all five Vanadium
d orbitals ($e_g$ and $t_{2g}$) are compared with the calculations performed with the ED impurity solver (filled symbols), which includes only the three $t_{2g}$ orbitals.}
\label{fig7}
\end{center}
\end{figure}

In this section, we extend the discussion of the single site DMFT results for \vox. 
We first discuss the Matsubara representation of the Green's function (the real and imaginary parts of the Green's function 
are shown respectively in Fig.~\ref{fig7}.\textbf{a} and Fig.~\ref{fig7}.\textbf{b}.)
We obtained, in particular, that the orbital 
density at the Fermi level 
$ n = -G'' \left( i \omega_n=0 \right) /  \pi $ 
obtained by CTQMC (open symbols)
is largest for the \dxy orbital. Interestingly, we find that the \eg states are almost empty and do not have a significant weight at the Fermi level.
The case for such an orbital selection process 
was argued long ago by Goodenough~\cite{vo2_paper_ref_goodenough}: \vo can be regarded as a 3d$^1$ system in the simple ionic picture
(with full charge transfer from the V 3d$^3$4s$^2$ to the two O 2p$^4$ configurations). 
The low energy states near the Fermi level are of strong vanadium 3d character and 
each vanadium atom is surrounded by an oxygen octahedron, such that the crystal field splits the 3d states into \tg and \eg bands. 
Since the structure is not cubic, however, 
the \tg states are further split into e$_{\pi g}$ and \ag states. 
In this simple ionic picture, the remaining 3d electron partially occupies the \ag 
band and the system remains metallic in the rutile phase. 
In the M$_1$ phase, pairs of vanadium atoms along the rutile axis form dimers. This Peierls distortion 
causes strong hybridization between the \ag orbitals of the two vanadium atoms, allowing the bonding state to fully fill and 
opening a gap between the bonding \ag and the unoccupied $\pi$ bands.

\begin{figure}
\begin{center}
\includegraphics[width=1.0\columnwidth]{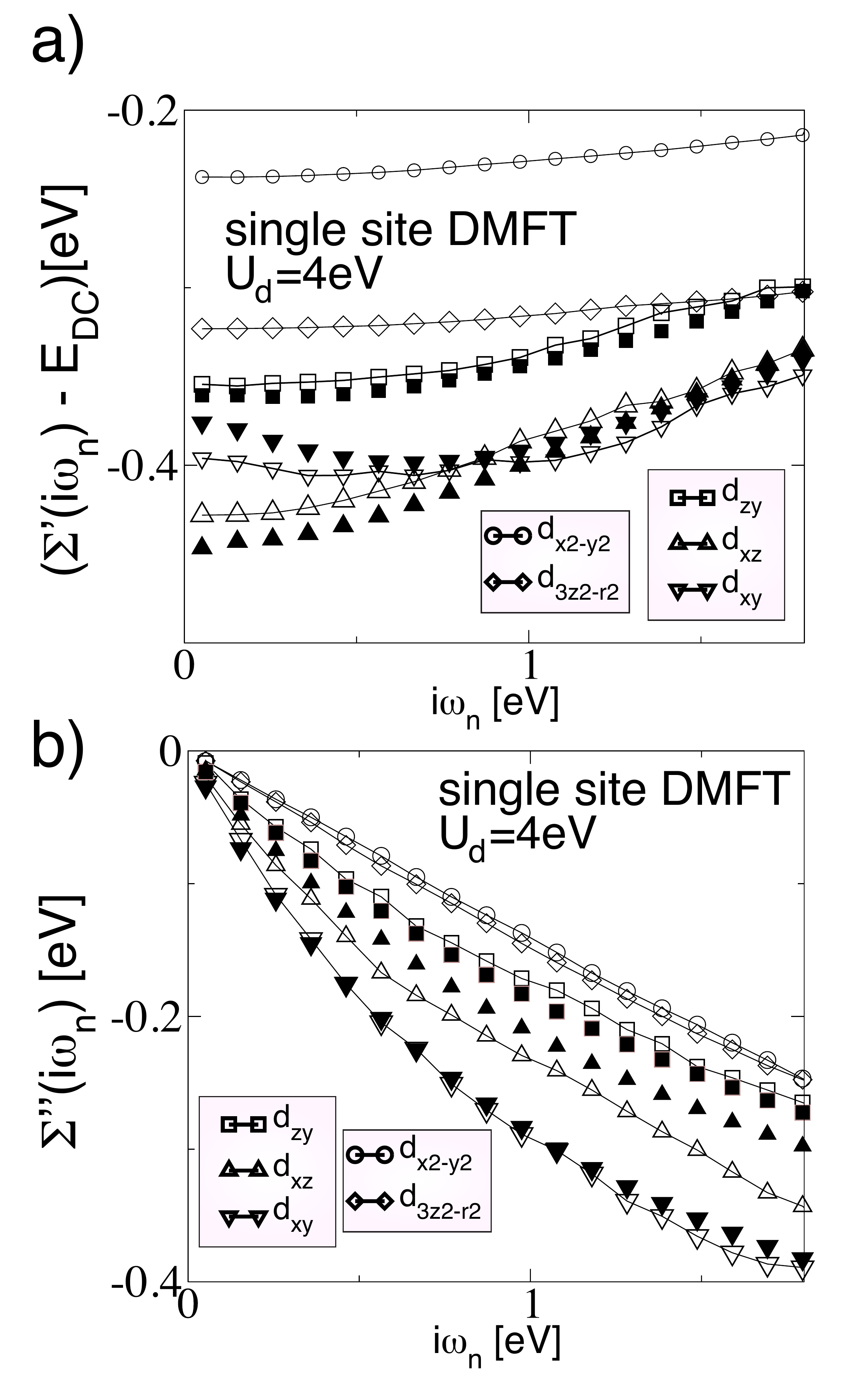}
\caption{ 
{\bf Matsubara representation of the self-energy:} (Color online) Real (a) and imaginary part (b) of the self-energy $\bold{ \tilde{\Sigma}}$
in the Matsubara representation for paramagnetic \vox. 
The calculations performed with the CTQMC 
impurity solver (open symbols) including all five Vanadium
d orbitals ($e_g$ and $t_{2g}$) are compared with the calculations performed using the ED impurity solver (filled symbols), which includes only 
the three $t_{2g}$ orbitals.}
\label{fig8}
\end{center}
\end{figure}

While the \tg orbitals point between the oxygen sites, the \dxd~and \dzd~orbitals point towards the oxygen sites, 
giving strong $\sigma$-bonds with the O 2p orbitals:
because of the high electronegativity of oxygen ($3.5$) compared to vanadium ($1.6$), 
the bonding \dzd~and \dxd~orbitals are of mainly O 2p character 
and hence the electrons are more localized to the O 2p than the V 3d orbitals; an almost ideal ionic bond. 
This suggests that the most localized orbitals in the 3d shell 
are the \tgd, and that polarized X-ray spectroscopy mainly resolves \tg orbitals at the vanadium sites. 

This scenario is also supported by the CTQMC data (Fig.~\ref{fig7}.\textbf{a,b}), which shows that dynamic correlations effects are not important
for the \eg orbitals. 
We obtained, from DFT+DMFT solved with the CTQMC, total densities of ($0.68$,$0.73$,$0.52$,$0.52$,$0.68$) respectively in the components of the 3d subshell (\dxd,\dzd,\dyzd,\dxzd,\dxyd).
Thus, there is $n=1.4$~e in the \eg and $1.7$~e in the \tg orbitals, 
in broad agreement with the $n=1.83$~e occupation of the 3d shell extracted from XAS 
measurements \cite{vo2_paper_ref2} and $n=1.9$~e obtained using photoemission spectroscopy \cite{vo2_paper_ref_9_photoemission}.

The real and imaginary parts of the self-energy are shown in Fig.~\ref{fig8}.\textbf{a} and Fig.~\ref{fig8}.\textbf{b}.
The dynamical correlations, described by the imaginary part of the self-energy, 
show that the \dxy orbital is also the most correlated orbital, whereas the \eg states are weakly correlated.
The latter is also supported by the weak frequency dependence of the \eg states in the real part of the self energy (Fig.~\ref{fig8}.\textbf{a}). 
Since the \eg orbitals do not exhibit any strong dynamical correlation, it is a good approximation to treat them at the Hartree level and thus we applied the ED solver only to the \tg orbitals.
We tested this approximation by comparing the densities obtained with the two methods. 
In particular, we obtain by using this approximation and the ED solver a total density in the \tg orbital $n=1.69$~e, in very good agreement with the CTQMC calculation, which finds a density of $n=1.7$~e in the \tg orbitals, 
and includes all the five d orbitals within the calculation. The comparison between the ED (filled symbols in Fig.~\ref{fig7}.\textbf{a,b} and Fig.~\ref{fig8}.\textbf{a,b}) and CTQMC (open symbols  in Fig.~\ref{fig7}.\textbf{a,b} and Fig.~\ref{fig8}.\textbf{a,b}) is remarkable for both the Green's function (Fig.~\ref{fig7}.\textbf{a,b}) and self-energy (Fig.~\ref{fig8}.\textbf{a,b}). 
Finally, we obtain also a very good agreement for the quasi-particle weight $Z_d$ obtained with the ED ($Z_d=0.77$) and CTQMC ($Z_d=0.73$) solvers.
The data shown in the paper and supplementary material were obtained with the ED solver when not specified otherwise. 


\section{Real space representation of the self-energy obtained by c-DMFT}

In this section, we extend the discussion regarding the cellular c-DMFT calculations under the presence of external
perturbations, such as disorder and vacancies.
\begin{figure}
\begin{center}
\includegraphics[width=1.0\columnwidth]{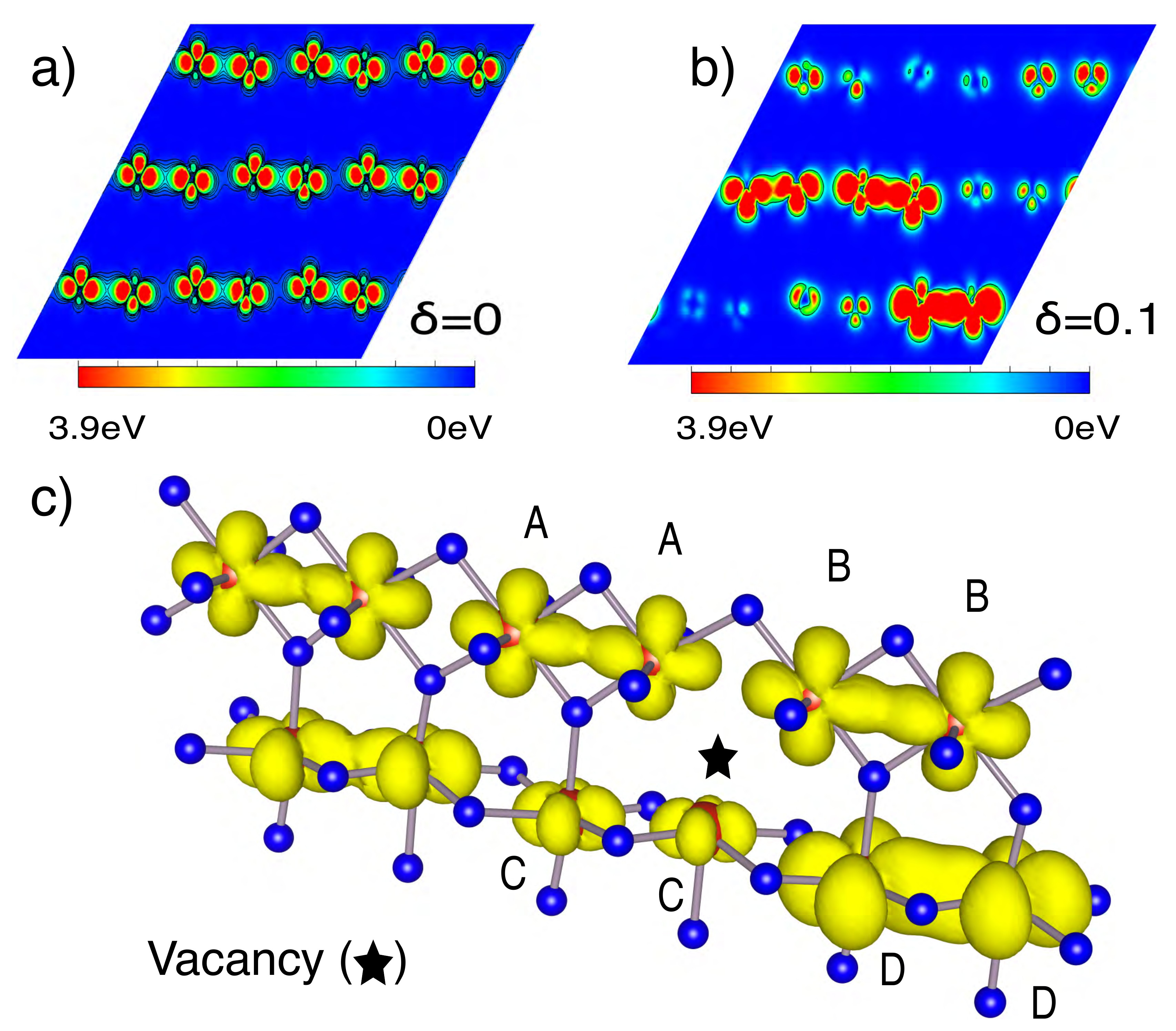}
\caption{
{\bf Real space representation of the scattering rate for cellular DMFT calculations: } (Color online) 
Contour plot of the real-space representation of the scattering rate at the Fermi level $-\Sigma'' \left( \mathbf{r},\mathbf{r} \right)$ obtained
by cellular DMFT (c-DMFT) shown for a $\left[ 010 \right]$ cut of the supercell-cell with zero (a) and finite disorder ($\delta=0.1$~\AA) (b).
c) Real-space iso-surfaces of the scattering rate at the Fermi level for the case of the O vacancy (star). The large (small) spheres denote
V (O) atoms. The rutile axis is oriented along the horizontal direction in all the calculations above.}
\label{fig6}
\end{center}
\end{figure}
Contour plots of the real-space 
representation of the local part of the scattering rate, 
$-\Sigma'' \left( \mathbf{r},\mathbf{r} \right)$, obtained
by cellular DMFT (c-DMFT), for a $\left[ 010 \right]$ 
cut of the super-cell, are shown for zero and finite disorder
in Fig.~\ref{fig6}.\textbf{a} and Fig.~\ref{fig6}.\textbf{b}, respectively.
Interestingly, we find that a moderate disorder of $\delta=0.1$~\AA~enhances the dimer bonding character by reducing the V-V distance among some of the
dimers and thus increasing the self-energy. This is illustrated by the real-space representation of the scattering rate,
in the $\left[ 010 \right]$ plane, of the super-cell for 
$\delta=0$~\AA~(Fig.~\ref{fig6}.\textbf{a}) and 
$\delta=0.1$~\AA~(Fig.~\ref{fig6}.\textbf{b}). 
Significantly, although 
the bonding character of some dimers is 
inevitably diminished, the integrated spectral function (Fig. 3.\textbf{a} in the paper) shows that the 
Mott localization effect is strong enough to sustain an insulator, 
such that no spectral weight is present at the Fermi level.

A very different effect is observed for the case of a single O vacancy.
At $\delta=0$~\AA, this defect does not affect the band edges but instead induces a mid-gap state with a finite
weight at the Fermi level. The real-space iso-surfaces of the scattering rate show (Fig.~\ref{fig6}.\textbf{c}) that the vacancy breaks a dimer
(dimer C) and enhances the bonding of the nearest neighbor dimer (D).
In particular, the strong mass renormalization associated with the Mott instability prevents the charge to flow freely 
between neighbour dimers, such that the mid-gap state remains essentially local in space.  
We argue that this mid-gap, localized state is relevant to the thin-film deposition of \vo layers,
where it was recently found that the strain of the substrate might induce subtle deviation from the 
canonical stoichiometry, and moreover 
that the metal-insulator transition was strongly varied over different samples~\cite{vo2_structure_power_is_better}.


%